\documentclass[draft]{agujournal}

\journalname{Earth and Space Science}
\usepackage{amsmath}
\usepackage{cases}
\usepackage{multirow}
\usepackage{graphicx}  

\begin{document}

%
%

\title{Machine Learning Approach for Solar Wind Categorization}

%
%

\authors{Hui Li \affil{1,4}, Chi Wang \affil{1,4}, Cui Tu \affil{2,4}, and Fei Xu \affil{3}}


\affiliation{1}{State Key Laboratory of Space Weather, National Space Science Center, CAS, Beijing, 100190, China.}
\affiliation{2}{State Key Laboratory of Space Weather, National Space Science Center, CAS, Beijing, 100190, China.}
\affiliation{3}{Physics Department, Nanjing University of Information Science and Technology, Nanjing, China.}
\affiliation{4}{University of Chinese Academy of Sciences, Beijing, 100049, China.}


\correspondingauthor{Hui Li}{hli@nssc.ac.cn}


\begin{keypoints}

\item An eight-dimensional scheme for 4-type solar wind categorization is developed based on 10 supervised machine-learning classifiers.
\item Machine learning approach significantly improves the classification accuracy by $\sim$ 10\% over existing manual schemes.
\item Classification only depends on typical solar wind observations, such as $N_\alpha, N_p, B_T, V_p, T_p$.
\end{keypoints}

%
%


\begin{abstract}

Solar wind classification is conducive to understand the physical processes ongoing at the Sun and solar wind evolution in the interplanetary space, and furthermore, it is helpful for early warning of space weather events. With rapid developments in the field of artificial intelligence, machine learning approaches are increasingly being used for pattern recognition. In this study, an approach from machine learning perspectives is developed to automatically classify the solar wind at 1 AU into four types: coronal-hole-origin, streamer-belt-origin, sector-reversal-region-origin, and ejecta. By exhaustive enumeration, an eight-dimensional scheme ($B_T$, $N_P$, $T_P$, $V_P$, $N_{\alpha p}$, $T_{exp}/T_P$, $S_p$, and $M_f$) is found to perform the best among 8191 combinations of 13 solar wind parameters. 10 popular supervised machine learning models, namely $k$ Nearest Neighbors (KNN), Support Vector Machines with linear and Radial Basic Function kernels, Decision Tree, Random Forest, Adaptive Boosting, Neural Network, Gaussian Naive Bayes, Quadratic Discriminant Analysis, and Extreme Gradient Boosting, are applied to the labeled solar wind data sets. Among them, KNN classifier obtains the highest overall classification accuracy, 92.8\%. It significantly improves the accuracy by 9.6\% over existing manual schemes. No solar wind composition measurements are needed, permitting our classification scheme to be applied to most solar wind spacecraft data. Besides, two application examples indicate that solar wind classification is helpful for the risk evaluation of predicted magnetic storms and surface charging of geosynchronous spacecrafts.

\end{abstract}

%
%

\section{Introduction} \label{sec:intro}

In 1959, the first solar wind observation was made by the Soviet satellite, \textit{Luna 1}. Since then, decades of in-situ solar wind measurements have firmly established that the solar wind plasma comes from different origins, for example, the coronal hole, the streamer belt, and active regions. \citet[][and references therein]{Xu and Borovsky 2015} showed that the solar wind can generally be classified into three major types: coronal-hole-origin plasma, streamer-belt-origin plasma, and ejecta.

Coronal-hole-origin plasma (CHOP) is sometimes called the fast solar wind, which originates from the open field line regions of coronal holes, and typically exhibits speeds in excess of 600 km/s at 1 AU and beyond \citep{Sheeley et al 1976,McComas et al 2008}. Statistically, CHOP tends to be homogeneous \citep{Bame et al 1977} with a high proton temperature and low plasma density \citep{Schwenn 2006}, and is dominated by outward propagating Alfv\'enic waves \citep{Luttrell and Richter 1988}. It exhibits a statistical non-adiabatic heating of the protons between 0.3 to 1.0 AU \citep{Hellinger et al 2011}. In addition, field-aligned relative drifts between the alphas and protons can frequently be found in CHOP, with a speed up to the local proton Alfv\'en speed \citep{Marsch et al 1982}. Moreover, the relative fluctuations of magnetic field and solar wind velocity are large in CHOP, about 24\% and 19\%, respectively. However, the corresponding Fourier spectral indices are -1.56 and -1.55 \citep{Borovsky 2012}, which is more likely to Iroshnikov-Kraichnan's theory ($f^{-3/2}$). As proposed by \citet{Li et al 2011}, this further indicates that current sheets are rare in such kind of solar wind.

Streamer-belt-origin plasma (SBOP), also known as the slow solar wind, has a typical speed less than 400 km/s. Compared to CHOP, SBOP does not exhibit much Alfv\'enic fluctuation \citep{Schwenn 1990} but is highly structured \citep{Bame et al 1977} with a low proton temperature and high plasma density \citep{Schwenn 2006}. In addition, the alpha-proton relative drift is typically absent in SBOP \citep{Asbridge et al 1976}, and the protons are closer to adiabatic \citep{Eyni and Steinitz 1978}. The relative fluctuations of magnetic field and solar wind velocity are small in SBOP, which are only 16\% and 11\%, respectively. Different from the situations in CHOP, both of the corresponding Fourier spectral indices obey Kolmogoroff's law ($f^{-5/3}$), giving -1.70 and -1.67, respectively \citep{Borovsky 2012}. This indicates that the solar wind may contains many current sheet structures \citep{Li et al 2011}.

Recently, it is found that SBOP can be further divided into two subgroups according to whether there exists an interplanetary magnetic sector reversal \citep{Antonucci et al 2005,Schwenn 2006}. One subgroup is referred to as streamer belt plasma (SBP) without sector reversals, and the other one is referred to as sector reversal region plasma (SRRP) with one sector reversal. The origin mechanism of SBP at the Sun is still a major unsolved problem in solar physics. There are two main mechanisms of SBP origination. One is the interchange magnetic reconnection of open field lines with closed streamer belt field lines \citep{Fisk et al 1999,Subramanian et al 2010,Antiochos et al 2011,Crooker et al 2012}; the other one is from the edge of a coronal hole near a streamer belt \citep{Wang and Sheeley 1990,Arge et al 2003}. SRRP is suggested to be emitted from the top of the helmet streamers \citep{Gosling et al 1981,Suess et al 2009,Foullon et al 2011}. Statistically, SBP and SRRP have different characteristics in the solar wind and subsequent effects on the geospace environments, which have been summarized by \citet{Borovsky and Denton 2013}.

Another major category of solar wind plasma is the so called ejecta (EJECT), which are associated with solar transients such as interplanetary coronal mass ejections (ICMEs) and magnetic clouds (MCs) \citep{Richardson et al 2000,Zhao et al 2009}. The origination of EJECT is the magnetic reconnection associated with the structures of streamer belts or active regions, which can impulsively emit plasma and make the magnetic field deviate from the Parker spiral \citep{Borovsky 2010}. The typical signatures of EJECT at 1 AU have been well summarized \citep[see][and references therein]{Zurbuchen and Richardson 2006}, for example, enhanced and smoothly rotating magnetic field, low proton temperature and plasma $\beta$, extreme density decrease, enhanced density ratio between alpha and proton, abundance and charge state anomalies of heavy ion species, bidirectional strahl electron beams, cosmic ray depletion, and declining velocity. Different from the expansions of CHOP or SBOP in the two directions transverse to radially outward from the Sun, impulsive EJECT expands in all three directions as they propagate outward \citep{Klein and Burlaga 1982}. Recently, \citet{Li et al 2016} performed a statistical survey on Alfv\'enic fluctuations inside ICMEs, finding that only 12.6\% of EJECT are found to be Alfv\'enic, and such a percentage decays linearly in general as the radial distance increases. The relative fluctuations of magnetic field and solar wind velocity are medium in EJECT, 21\% and 15\%, respectively \citep{Borovsky 2012}. The Fourier spectral indices are close to -5/3 \citep{Borovsky 2012}, and may decrease as the radial distance increases \citep{Li et al 2017}.

The categorization of the solar wind into its origin is of great importance for solar and heliospheric physics studies. Firstly, the statistical properties of solar wind should be clarified by its type to make a more comprehensive understanding of solar wind. Secondly, dividing the solar wind observations at 1 AU according to their origins can lead to a better diagnosing of physical processes ongoing at the Sun \citep{Mariani et al 1983,Thieme et al 1989,Thieme et al 1990,Matthaeus et al 2007,Borovsky 2008,Zastenker et al 2014}. Thirdly, the geoeffectiveness (geomagnetic activity, specifically, magnetic storm and substorm) of solar wind from different origins differ considerably \citep[e.g.,][]{Borovsky and Denton 2006,Turner et al 2009,Borovsky and Denton 2013}. Such a categorization would be helpful for the early warnings of space weather. Note that, these differences are in statistical terms. For individual cases, the situations may be quite different and complicated.

Usually, the solar wind classification is done manually by experienced people. In the literature, several empirical categorization methodologies in different parameter space have been proposed. In a one-dimensional parameter space, the solar wind was usually separated into ``fast wind" or ``slow wind" according to its speed, $V_p$ \citep{Arya and Freeman 1991,Tu and Marsch 1995,Feldman et al 2005,Yordanova et al 2009}. However, such a $V_p$ scheme can only roughly divide the solar wind into CHOP and SBOP, but could not separate out EJECT, SBP, and SRRP. Moreover, the criterion of $V_p$ is not unique. In 2014, another one-dimensional scheme based on the parameter $P_{type}$ (= $2\log S_p-\log(C^{6+}\/C^{5+})-\log(C^{7+}\/C^{6+}))$ was proposed by \citet{Borovsky and Denton 2014}. As the understanding of ICMEs and MCs is getting better, many methodologies have been proposed to identify EJECT \citep[see][and references therein]{Zurbuchen and Richardson 2006,Kunow et al 2006}, and several catalogs of EJECT at 1 AU have been produced \citep[e.g.,][]{Lepping et al 2005,Jian et al 2006,Richardson and Cane 2010}. Recently, the composition measurements were used for solar wind classification. An algorithm in a two-dimensional parameter space, such as O$^{7+}$/O$^{6+}$ and $V_p$, was constructed by \citet{Zurbuchen et al 2002,Zhao et al 2009,von Steiger et al 2010}. Such a two-dimensional scheme is still not able to divide SBOP into SBP and SRRP. In addition, such a scheme is not generally available for most solar wind spacecrafts due to the lack of on-board ion composition instruments. \citet{Xu and Borovsky 2015} developed a three-parameter, four-plasma-type categorization scheme based on commonly used solar wind measurements, and obtained a good classification accuracy. In addition, an on-board solar wind classification algorithm was already applied in the \textit{Genesis} spacecraft \citep{Neugebauer et al 2003,Reisenfeld et al 2003}. Such a automatic method requires the measurement of bi-directional electron and historic solar wind classification results.

Although the traditional classification has significant improvements in recent decades, there remains some improvement room for the existing empirical categorization schemes. The multi-label classification is regarded as a typical task of machine learning. Recently, the performance of machine learning classification is getting much better as the rapid developments of artificial intelligence theory and techniques. Machine learning technique is becoming more and more popular and powerful in big-volume data analysis in space physics, which may offer a solution to improve the accuracy of solar wind classification. \citet{Camporeale et al 2017} recently employed a machine learning technique, Gaussian Process, in a four-category solar wind classification, and obtained a median accuracy larger than 96\% for all categories. However, the time resolutions of the variables they used are not uniform. For example, the temporal resolution is one day for sunspot number and solar radio flux (10.7 cm), but is one hour for the other five solar wind parameters and for the reference solar wind data sets. \citet{Camporeale et al 2017} did not demonstrate the reasonableness of such mixture of hourly averaged solar wind parameters and daily sampled parameters.

In this work, we will apply 10 popular supervised machine learning models to identify the solar wind plasma into four types (CHOP, SBP, SRRP, EJECT) based on typical solar wind observations with the same temporal resolution of 1 hour. In particular, we will identify the best parameter scheme from 8191 combinations of 13 parameters derived from typical solar wind observations, judged by the classification accuracy as high as possible.

\section{Methodology} \label{sec:met}
For conventional classifications of the solar wind plasma at 1 AU, reference solar wind data with known plasma types should be first collected. Then, empirical relationships are developed to describe the domains of different plasma in some parameter space. Generally, the human experience performs well in two/three-dimensional parameter space. For a multi-dimensional space, humans cannot easily derive the empirical relationships.

For supervised machine learning approaches, reference solar wind data with known plasma types are needed for training the classifier as well. Then, the discriminant rules would be developed automatically by machine learning classifiers. One of the advantages is that the discriminant rules can be easily obtained in a multi-dimensional space for the machine learning perspective. Usually, 75\% (80\%) of the reference solar wind data are used for training, and the remaining 25\% (20\%) used for testing, especially for the situation with the cases less than 10000.

\begin{table}[htbp]
\renewcommand{\thetable}{\arabic{table}}
\caption{10 machine learning classifiers used in this study.}
\centering
\begin{tabular}{c c c}
\hline
\hline
 Classifier & Abbreviation & Reference \\
\hline
  $k-$nearest neighbors & KNN & \citet{Denoeux 1995} \\
  linear support vector machine & LSVM & \citet{Fan et al 2008} \\
  SVM with Gaussian radial basis function kernel & RBFSVM & \citet{Buhmann 2003} \\
  Decision Trees & DT & \citet{Breiman et al 1984} \\
  Random Forests & RF & \citet{Ho 1995} \\
  Adaptive Boosting & AdaBoost & \citet{Zhu et al 2009} \\
  Neural Network & NN & \citet{Rojas 1996} \\
  Gaussian Naive Bayes & GNB & \citet{Perez et al 2006} \\
  Quadratic Discriminant Analysis & QDA & \citet{Srivastava et al 2007} \\
  Extreme Gradient Boosting & XGBoost & \citet{Chen and Guestrin 2016} \\
\hline
\hline
\end{tabular}
\label{models}
\end{table}

\subsection{Machine Learning Classifiers}
Classification is regarded as one of the typical tasks carried out by so-called machine learning system. The classifier is a critically important part of machine learning toolkit. As the rapid development of machine learning technique, a large number of classification algorithms have been developed. In this study, we will apply 10 widely used classifiers \citep{Cady 2017} to perform solar wind categorization, namely $k-$nearest neighbors (KNN), linear support vector machines (LSVM), SVM with a kernel of Gaussian radial basis function (RBFSVM), Decision Trees (DT), Random Forests (RF), Adaptive Boosting (AdaBoost), Neural Network (NN), Gaussian Naive Bayes (GNB), Quadratic Discriminant Analysis (QDA), and eXtreme Gradient Boosting (XGBoost). Table \ref{models} gives the references of these 10 classifiers for readers to get more details. All the classification algorithms are included in the Scikit-learn package, which is an open source machine learning library written in the Python programming language \citep{Pedregosa et al 2011}. In this work, we will use the Scikit-learn package to carry out solar wind classifications. The details of the Scikit-learn package can be found at http://scikit-learn.org/stable/index.html.

\subsection{Reference Solar Wind data}
For supervised machine learning, reference solar wind data sets with known types are needed to train the classifiers. We use the same data sets utilized in \citep{Xu and Borovsky 2015}, and the solar wind plasma will be divided into four types: CHOP, SBP, SRRP and EJECT. The collection of reference CHOP comes from the ideal events used by \citet{Xu and Borovsky 2015}. They examined the solar wind speed $V_p$, the proton-specific entropy $S_p = T_p/N_p^{2/3}$, O$^{7+}$/O$^{6+}$, C$^{6+}$/C$^{5+}$ and the characteristics of the interplanetary magnetic field to identify CHOP. The intervals of twenty-seven day repeating steady high-speed solar wind streams with long intervals (days) are regarded as CHOP. CHOP starts after the compression of the corotating interaction region (CIR) and ends before the onset of the trailing edge rarefaction. At the same time, they also excluded large jumps in $S_p$, O$^{7+}$/O$^{6+}$ or C$^{6+}$/C$^{5+}$ to make sure CHOP were not contaminated with ejecta. A total of 3049 hours of CHOP identified by \citet{Xu and Borovsky 2015} are used here.

The collection of reference SBP comes from the pseudo-streamers during 2002-2008 identified by \citet{Borovsky and Denton 2013}. Looking earlier in time the plasma upstream of the CIR, they checked the preceding intervals of CHOP. If the preceding coronal hole was of the same magnetic sector as the coronal hole immediately following the CIR, and if no sector reversals occurred in the streamer belt origin plasma between the successive two coronal holes, then the streamer belt plasma was classified into SBP. A total of 2275 hours of SBP identified by \citet{Borovsky and Denton 2013} are used here.

The collection of reference SRRP also comes from the work done by \citet{Xu and Borovsky 2015}. They examined the electron strahl observation and found some broad regions where the electron strahl dropped out around magnetic sector reversals at 1 AU. They denoted the regions where the strahl was very weak, intermittent, and/or intermittently bi-direction just outside the strahl dropped out regions, to be ``strahl confusion zones''. The solar wind from these confusion zones are defined as SRRP. A total of 1740 hours of SRRP are used here.

The magnetic cloud collection made by \citep{Lepping et al 2005} is used to represent EJECT here, which can be found at \textit{http://wind.gsfc.nasa.gov/mfi/mag\_cloud\_pub1.html}. Magnetic clouds are believed to be a subset of interplanetary coronal mass ejections (ICMEs) with an enhancement of magnetic field intensity and a gradual rotation in direction. The typical properties of magnetic cloud are a flux rope field configuration, low proton temperatures and low plasma beta value \citep{Klein and Burlaga 1982}. In general, only about one third of ICMEs can be regarded as magnetic clouds \citep{Bothmer and Schwenn 1996,Richardson and Cane 2004}. \citet{Xu and Borovsky 2015} found a dual-population structure for the collection of ICMEs identified by \citet{Richardson and Cane 2010}, but a single population for the collection of magnetic clouds identified by \citet{Lepping et al 2005}. They believed that magnetic clouds can better present ejecta from the Sun, while the collection of ICMEs probably contains some non-ejecta data. A total of 1926 hours of EJECT are used here.

\begin{table}[htbp]
\renewcommand{\thetable}{\arabic{table}}
\caption{List of 13 parameters used for solar wind classification.}
\centering
\begin{tabular}{c c }
\hline
\hline
 Parameter & Symbol \\
\hline
  magnetic filed intensity & $B_T$ \\
  proton density & $N_p$  \\
  proton temperature & $T_p$ \\
  solar wind speed & $V_p$ \\
  proton-specific entropy & $S_p$ \\
  Alfve\'n speed & $V_A$  \\
  temperature ratio & $T_{exp}/T_p$ \\
  ratio of proton and alpha number density & $N_{\alpha p}$ \\
  dynamic pressure & $P_d$ \\
  solar wind electric field &  $E_y$ \\
  plasma beta value & $\beta$ \\
  Alfv\'en Mach number & $M_A$ \\
  fast magnetosonic Mach number & $M_f$ \\
\hline
\hline
\end{tabular}
\label{parameters}
\end{table}

After removing some data gaps, the reference data set is composed of 2881 (33.4\%) 1-hr events categorized as CHOP, 2215 (25.7\%) events of SBP, 1694 (19.6\%) events of SRRP, and 1835 (21.3\%) events of EJECT. The imbalance ratio of these four types solar wind may affect the classification accuracy. In general, the accuracy would be relatively low when fewer reference solar wind are used for training. The ratio of reference SRRP is the lowest. Its classification accuracy is indeed found to be lower than the other three types in the following section. The solar wind parameters used in this study are from the OMNI database (http://omni.gsfc.nasa.gov/), which is primarily a 1963-to-current compilation of hourly-averaged, near-Earth solar wind magnetic field and plasma parameter data from several spacecrafts in geocentric or L1 (Lagrange point) orbits. The data have been extensively cross compared and cross-normalized for some spacecrafts and parameters.

\begin{figure}[htbp]
\centering
\noindent\includegraphics[width=32pc]{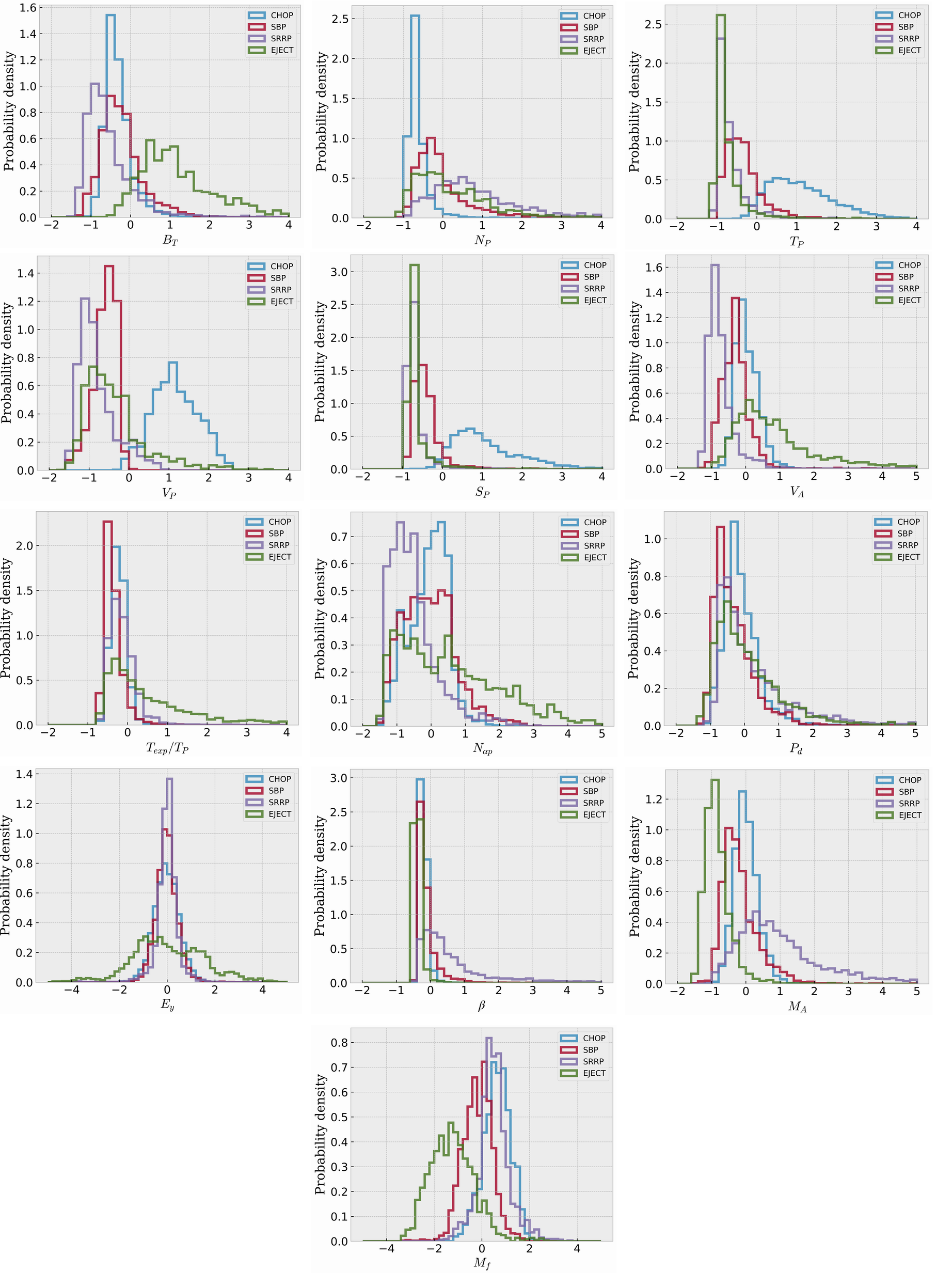}
\caption{Probability density distributions of 13 solar wind parameters calculated from the whole reference solar wind data sets. The parameters have been rescaled as follows: $X = (X - \overline{X}) / \sigma_X$. The area under each curve equals 1. }
\label{pdf}
\end{figure}

\section{Categorization Results}
With the input of solar wind parameters and information of solar wind types, the classifiers can build discriminant rules automatically based on machine learning algorithms. Note that, most solar wind spacecrafts have no composition instrumentation. To make the applicability of our classification scheme more extensive, the typical solar wind observations (the magnetic field intensity, $B_T$, the proton number density, $N_p$, the alpha particle number density, $N_{\alpha}$, the proton temperature, $T_p$, and the solar wind speed, $V_p$) and their derived quantities are used here. As listed in Table \ref{parameters}, a total of 13 parameters are used for solar wind classification, such as $B_T$, $N_p$, $T_p$, $V_p$, the proton-specific entropy, $S_p$, the Alfve\'n speed, $V_A=B_T/\sqrt{\mu_0m_pN_p}$ ($\mu_0$ is the permeability in vacuum and $m_p$ is the mass of proton), the temperature ratio, $T_{exp}/T_p$ ($T_{exp}=(V_p/258)^{3.113}$ is the velocity-dependent expected proton temperature given by \citet{Xu and Borovsky 2015} in unit of eV), the number density ratio of proton and alpha, $N_{\alpha p}$, the dynamic pressure, $P_d$, the solar wind electric field, $E_y$, the plasma beta value $\beta$, the Alfv\'en Mach number, $M_A=V_A/V_p$, and the fast magneto-sonic Mach number, $M_f=V_p/\sqrt{C_s^2+V_A^2}$ ($C_s$ is the acoustic velocity). Note that, this parameter list includes all the parameters used in \citep{Xu and Borovsky 2015} and four of seven parameters used in \citep{Camporeale et al 2017}. As mentioned, the reference solar wind with known types is from hourly-averaged OMNI database, thus, only the parameters with a temporal resolution of one hour are considered here. The parameters with a temporal resolution of one day used in \citep{Camporeale et al 2017}, the Sunspot number and solar radio flux (10.7 cm), are not considered here. Among them, a specific combination of parameter with the highest classification accuracy will be chosen for further analysis.

Figure \ref{pdf} shows the probability density distributions of the above 13 parameters calculated from the whole reference solar wind data sets. Similar probability density distributions of $V_p$, $V_A$, $S_p$, and $T_{exp}/T_p$, are also shown by \citet{Camporeale et al 2017}. Note that, the parameters have been rescaled as follows: $X = (X - \overline{X}) / \sigma_X$, where $\overline{X}$ represents the mean value of a parameter, and $\sigma_X$ denotes the standard deviation. Obviously, it is difficult to distinguish the 4-type solar wind well from any individual probability distribution, which motivates the classification in a multi-dimensional space. Nevertheless, some parameters could contribute to distinguish some solar wind type from the others. For example, $B_T$ and $M_f$ contribute to distinguish EJECT from the others, especially from the SRRP; $N_p$, $V_p$, and $N_{\alpha p}$ are useful to distinguish between CHOP and SRRP; $T_p$ and $S_p$ help to distinguish CHOP from the others; and $V_A$ is helpful to distinguish SRRP from the others. A natural thought is that the classification accuracy would be improved greatly by considering the above eight parameters together. Actually, the selected eight-dimensional parameter scheme with the best classification accuracy for KNN classifier contains 7 of the above 8 parameters, only with $V_A$ replaced by $T_{exp}/T_p$.

\begin{table}[htbp]
\renewcommand{\thetable}{\arabic{table}}
\caption{Classification performances for 10 classifiers based on the combination of $B_T$, $N_p$, $T_p$, $V_p$, $N_{\alpha p}$, $T_{exp}/T_p$, $S_p$, and $M_f$. From second to sixth column, the value gives the classification accuracy. The last column gives the Hanssen and Kuipers' Discriminant, HKSS. Note that, 75\% of the reference solar wind data are used for training, and the remaining 25\% used for testing. 100 iterations with random selection of the training data are run and the mean accuracies are reported here.}
\centering
\begin{tabular}{c c c c c c c c}
\hline
\hline
          & CHOP & SBP & SRRP & EJECT & 4-type & HKSS\\
\hline
    KNN   & 99.2 & 91.1 & 83.8 & 92.9 & 92.8 & 0.902 \\
  XGBoost & 99.2 & 90.9 & 83.6 & 92.8 & 92.6 & 0.898 \\
   RF     & 99.3 & 90.2 & 81.6 & 94.1 & 92.3 & 0.895 \\
   RBFSVM & 99.1 & 89.0 & 81.1 & 94.1 & 91.9 & 0.890 \\
   NN     & 99.1 & 88.7 & 80.6 & 92.2 & 91.3 & 0.881 \\
   DT     & 98.1 & 84.8 & 77.6 & 89.0 & 88.7 & 0.846 \\
   LSVM   & 99.0 & 81.1 & 71.1 & 88.2 & 86.6 & 0.816 \\
   QDA    & 98.7 & 80.4 & 75.0 & 73.7 & 84.0 & 0.779 \\
   GNB    & 96.8 & 76.0 & 76.9 & 73.1 & 82.5 & 0.767 \\
 AdaBoost & 97.5 & 85.1 & 45.2 & 85.6 & 81.5 & 0.737 \\
\hline
\hline
\end{tabular}
\label{perform}
\end{table}

Given 13 input features, a total of 8191 combinations exist. Taking the KNN classifier as an example, the classification accuracy is calculated by using all the 8191 combinations of input features. The eight-dimensional scheme, the combination of $B_T$, $N_P$, $T_P$, $V_P$, $N_{\alpha p}$, $T_{exp}/T_P$, $S_p$, and $M_f$ is found to perform the best, with the overall accuracy of 92.8\%. The accuracy for classifying CHOP, SBP, SRRP, and EJECT is 99.2\%, 91.1\%, 83.8\%, and 92.9\%, respectively. Although this scheme is choosing from 8191 combinations of 13 variables from the perspective of practical effect, it really has physical meanings. As shown in Figure \ref{pdf}, these parameters indeed contribute to distinguish some solar wind type from the others. If some new variables are considered, another method to determine the variable combination may also work and reduce the test number greatly. For example, identify the first variable, by using that alone the best classification accuracy can be obtained. Then, identify the second variable, by considering that with the first determined variable together the best classification accuracy can be obtained. At last, repeat the second step until the accuracy could not be improved by adding any new variable. Actually, a set of mutually independent variables contain enough information of the classification system. Here, some combined parameters, e.g., $S_p$, $V_A$, $T_{exp}/T_p$, etc, are used only for the perspective of improving the classification accuracy. If the mutually independent variables ($B_T$-$V_p$-$N_p$-$T_p$-$N_{\alpha p}$) are used, the classification accuracy of 4-type solar wind will decline slightly from 92.8\% to 92.0\%.

The classification is also done for the other 9 classifiers with the same parameter scheme used. The results are listed in Table \ref{perform}. Five classifiers, KNN, XGBoost, RF, RBFSVM, and NN, produce the accuracy better than 90\%. DT and LSVM also perform well, with the overall accuracy better than 85\%. The remaining classifiers, QDA, GNB, and AdaBoost, yield accuracies between 80-85\%. It should be mentioned that the overall accuracy for the other 9 classifiers should be improved if some special kind of parameter combination were used. The identification of CHOP is relatively easy. All the 10 classifiers work very well, with the accuracy better than 96.5\% and the highest accuracy given by RF of 99.3\%. For identifying EJECT, the accuracy decreases slightly. Only 5 classifiers yield accuracies better than 92\%, and the highest accuracy given by RBFSVM is 94.1\%. For identifying SBP, only 3 classifiers yield accuracies better than 90\%, with the highest accuracy given by KNN of 91.1\%. The identification of SRRP is relatively difficult. Only 5 classifiers yield accuracies better than 80\%, and the highest accuracy given by KNN is only 83.8\%. Note that, 75\% of the reference solar wind data are used for training, and the remaining 25\% used for testing. To make sure that our results are independent on the choice of training data set, cross validation is quite necessary. Thus, we perform 100 runs with the training data set being chosen randomly. The accuracy given in Table \ref{perform} is the averaged value of the 100 tests.

Besides the classification accuracy, the Hanssen and Kuipers' Discriminant, HKSS, is also given in Table \ref{perform}. The HKSS, also known as the True Skill Statistic, represents the classification accuracy relative to that of random chance. For multi-category classification, its expression can be written as follows:
\begin{linenomath}
\begin{equation}
   HKSS = \frac{\frac{1}{N}\sum_{i=1}^{K} n(F_i,O_i) - \frac{1}{N^2} \sum_{i=1}^{K} N(F_i)N(O_i)}{1-\sum_{i=1}^{K}(N(O_i))^2}
\end{equation}
\end{linenomath}
where $n(F_i,O_i)$ denotes the number of classifications in category $i$ that had observations in category $i$, $N(F_i)$ denotes the total number of classifications in category $i$, $N(O_i)$ denotes the total number of observations in category $i$, and $N$ is the total number of classification. HKSS ranges from -1 to 1. 1 represents the perfect performance, 0 denotes no improvement over a reference classification, and $\le 0$ indicates worse than the reference. From Table \ref{perform}, it is clear that the results of HKSS for 10 classifiers are similar with the results of accuracy.

To test the sensitivity of variable in our eight-dimensional scheme, one variable is in turn left out from the scheme and the accuracies are recalculated accordingly. When $S_p$ is not considered, the classification accuracy has the least decrease, 0.1\%. And the accuracy has the largest decrease, 2.2\%, when $N_{\alpha p}$ is not considered. However, it does not imply that $S_p$ is the least importance variable in solar wind classification. Actually, the highest classification accuracy is obtained by using $S_p$ alone, among the 13 variables. For different parameter combination, the most sensitive parameter should be different as well.

\begin{figure}[htbp]
\centering
\noindent\includegraphics[width=32pc]{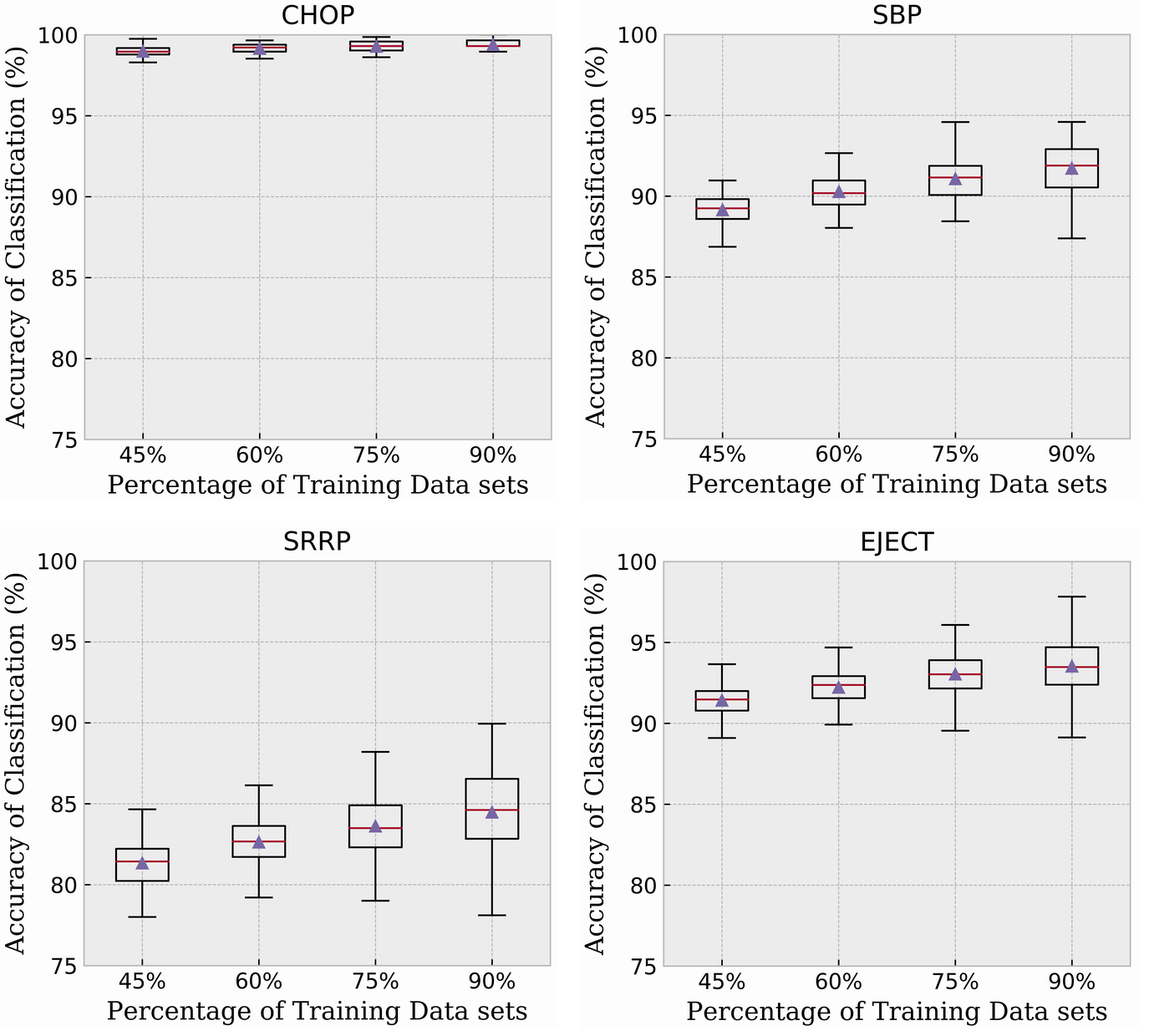}
\caption{Accuracy of the KNN classifier calculated from 100 runs with different ratio of training data set being chosen randomly. The boxes denote the first and third quartiles of the accuracy distribution. The horizontal lines and triangles represent the median and mean values, respectively. The whiskers denote the 2nd and 98th percentiles.}
\label{test}
\end{figure}

It is hard to make sure that the result of supervised machine learning is neither over-fitted nor under-fitted. Comparing the accuracy of training vs. testing data sets is a good way, but not sufficient. Cross validation is another strategy to overcome such problems. Following the methodology of \citet{Camporeale et al 2017}, we also compare the results of 100 runs for different ratios of the training data. In general, over-fitting is especially likely in cases where training examples are rare. Thus, a relative large ratio of training data, for example, 45\%, 60\%, 75\%, and 90\% are used, and the results are shown in Figure \ref{test}. The boxes denote the first and third quartiles of the accuracy distribution. The horizontal lines and triangles represent the median and mean values, respectively. The whiskers denote the 2nd and 98th percentiles. It is clear that the mean accuracy slightly increases when the ratio of training data increases from 45\% to 75\%. For the ratio of 90\%, the accuracy has no significant improvement, however, the variation amplitude of classification accuracy increases significantly, and the lowest accuracy even decreases slightly for identifying SBP, SRRP, and EJECT. In the following texts, the accuracies are all obtained by using 75\% of the data for training. This is just a simple approach to judge whether an over-fitting occurs or not. There may exist other, more robust, means of examining over-fitting or under-fitting. \citet{Camporeale et al 2017} showed the accuracy of the Gaussian Process classification model with 10\%, 15\%, 20\%, and 25\% of the original data used for training. Similarly, the accuracy increases when more data is used for training.

\begin{figure}[htbp]
\centering
\noindent\includegraphics[width=20pc]{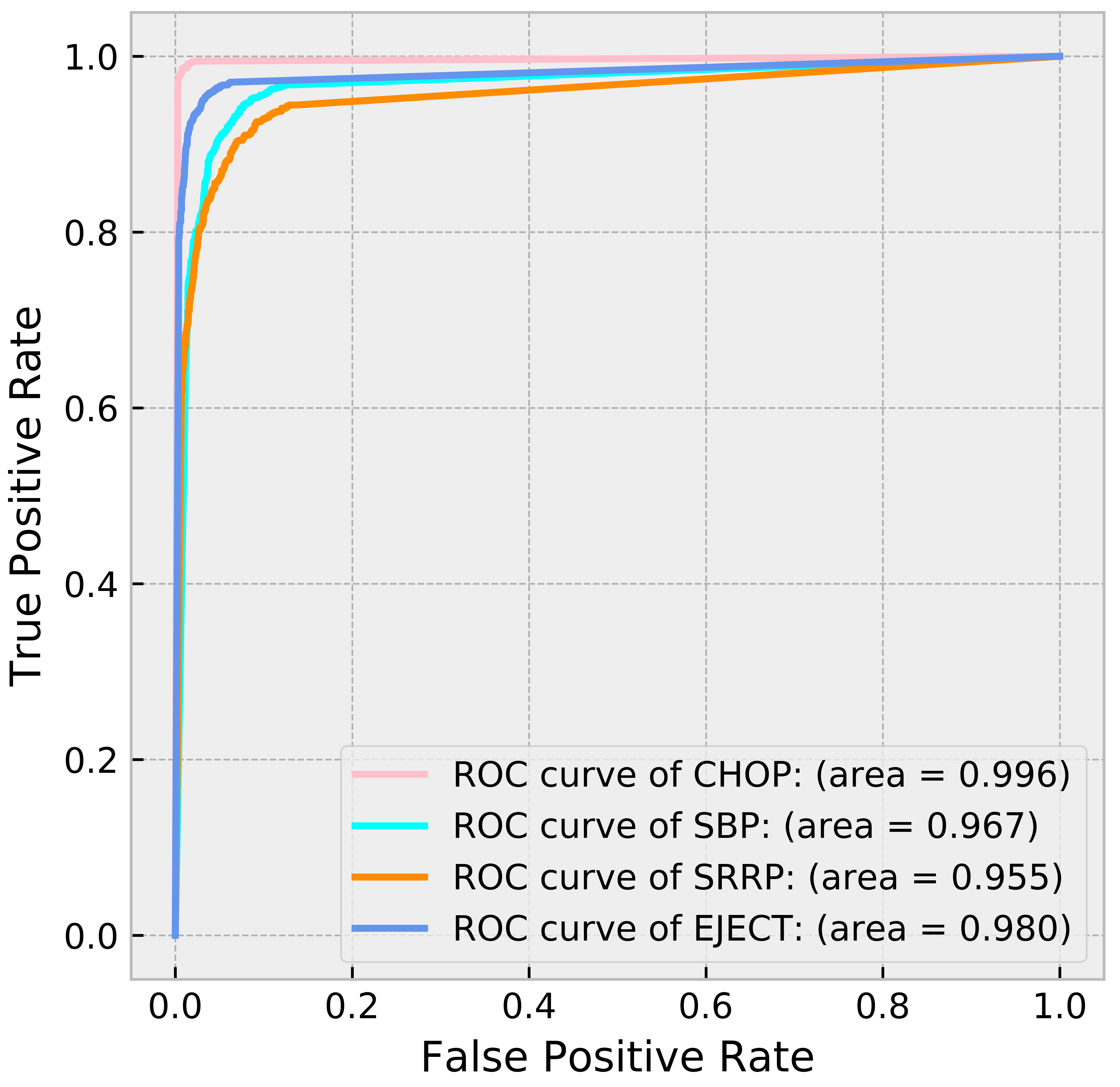}
\caption{Receiver operating characteristic (ROC) curves for CHOP, SBP, SRRP, and EJECT. The False Positive Rate is defined as the ratio of false positives divided by the total number of negatives. The True Positive Rate denotes the ratio of true positives divided by the total numbers of positives. The area of the curve represents the goodness of binary classification, and unity denotes the perfect result.}
\label{ROC}
\end{figure}

For binary classification, the threshold of probability changes to accuracy in terms of true and false positives and negatives. Here, ``true/false'' denotes correct, or incorrect, classification, and ``positive/negative'' denotes that the solar wind is classified to be, not to be, some type. Thus, ``true positive/flase positive'' denotes that the solar wind is correctly/incorrectly classified to be some type, while ``true negative/flase negative'' denotes that the solar wind is correctly/incorrectly classified not to be some type. The Receiver operating characteristic (ROC) curve for different values of thresholds gives a concise representation of this metric. The horizontal axis is the False Positive Rate (FTR), which is defined as the ratio of false positives divided by the total number of negatives. And the vertical axis is the True Positive Rate (TPR), which denotes the ratio of true positives divided by the total numbers of positives. A perfect classification would give FPR = 0, TPR = 1, and the area of ROC curve equals unity. Figure \ref{ROC} shows the ROC curves for CHOP, SBP, SRRP, and EJECT. The areas of the curves are 0.996, 0.967, 0.955, and 0.980, respectively, indicating that the classification is pretty good. From practice, the threshold of probability can be chosen to be 0.3-0.5 to obtain optimal FPR and TPR, which is consistent with \citet{Camporeale et al 2017}.

\begin{figure}[htbp]
\centering
\noindent\includegraphics[width=30pc]{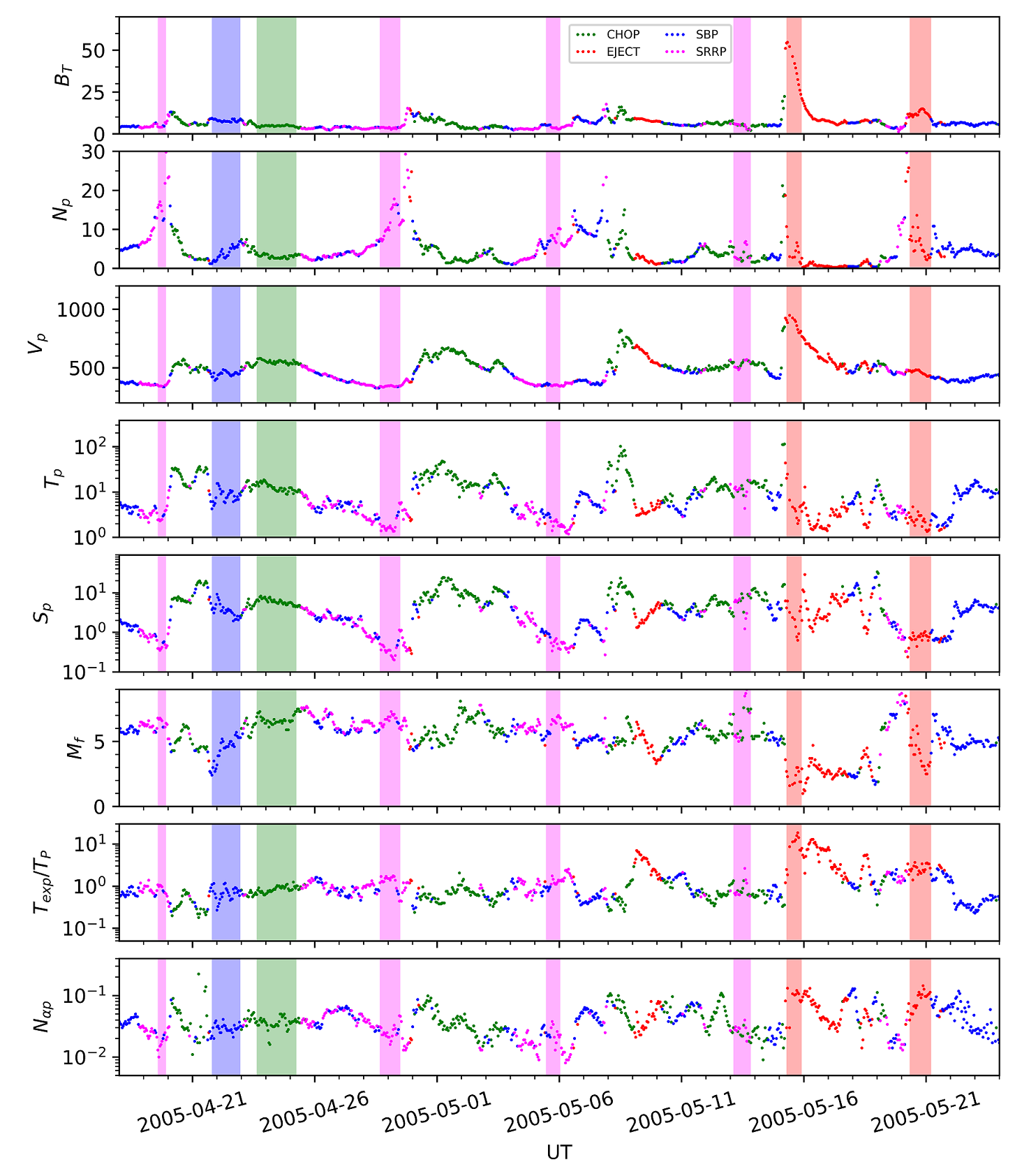}
\caption{An example of solar wind classification obtained by the KNN classifier. From top to bottom, the panel represents the magnetic field intensity, the proton number density, the solar wind speed, the proton temperature, the proton-specific entropy, the plasma beta value, the fast magneto-sonic Mach number, the dynamic pressure, and the ratio of proton and alpha number density. The units are in nT, cm$^{-3}$, km/s, eV, eVcm$^2$, unity, unity, nPa, and unity, respectively. The shaded regions represent the time intervals of reference solar wind with known types.}
\label{example}
\end{figure}

Figure \ref{example} shows an example of solar wind classification obtained by the KNN classifier. The shaded regions represent the time intervals of reference solar wind with known types. In general, all the solar wind can be distinguished well. It is clear that the CHOP, SBP, and EJECT in the shaded regions are identified perfectly with the accuracy nearly 100\%. The classification accuracy for SRRP is not so high but still good, $\sim$ 85\%. Occasionally, it is wrongly identified as SBP (on April 19, 28-29, and May 5) or CHOP (on May 13). Two long-duration EJECTs are also identified after CHOPs, for example, the EJECT on May 15-17, and 20, which had already been identified as two magnetic clouds by \citep{Lepping et al 2005}. At the same time, some short-duration EJECTs (several hours) are also identified on May 09-10 and 30-31, which may be the so-called small flux ropes proposed by \citet{Moldwin et al 2000}, and are in agreement with the small-scale magnetic flux rope database (\textit{http://fluxrope.info/}) given by Dr. Jinlei Zheng and Dr. Qiang Hu at University of Alabama Huntsville. This indicates that our categorization scheme may in certain cases be useful for identifying small flux ropes, but more investigation and validation is needed.

\begin{table}[htbp]
\renewcommand{\thetable}{\arabic{table}}
\caption{Accuracies of various categorization schemes in solar wind classification. Note that, 25\% of the database is used for training in \citep{Camporeale et al 2017}, but the ratio is 75\% in our study. 100 iterations with random selection of the training data are run and the mean accuracies are reported here.}
\centering
\begin{tabular}{c c c c c c c c}
\hline
\hline
 Accuracy (\%) & CHOP & SBP & SRRP & EJECT & 4-type \\
\hline
$O^{7+}/O^{6+}$-$V_p$ & \multirow{2}{*}{98.0} & \multirow{2}{*}{73.0} &  & \multirow{2}{*}{63.5} &  \\
 Zhao et al. [2009] &  &  &  &  &  &  &  \\ \hline
$S_p$-$V_A$-$T_{ex}/T_p$ & \multirow{2}{*}{96.9} & \multirow{2}{*}{69.9}  & \multirow{2}{*}{72.0} & \multirow{2}{*}{87.5} & \multirow{2}{*}{83.2} \\
Xu \& Borovsky [2015] &  &  &  &  &  &  &  \\ \hline
$S_p$-$V_A$-$T_{exp}/T_p$ & \multirow{2}{*}{97.2} & \multirow{2}{*}{74.9} & \multirow{2}{*}{69.7} & \multirow{2}{*}{88.7} & \multirow{2}{*}{84.3} \\
KNN (this work) &  &  &  &  &  &  &  \\ \hline
$V_p$-$\sigma_T$-$SSN$-$F10.7$-$V_A$-$S_p$-$T_{exp}/T_p$ & \multirow{2}{*}{99.7} & \multirow{2}{*}{98.7}  & \multirow{2}{*}{97.5} & \multirow{2}{*}{96.1} & \multirow{2}{*}{98.2} \\
Camporeale et al. [2017] &  &  &  &  &  &  &  \\ \hline
$V_p$-$\sigma_T$-$SSN$-$F10.7$-$V_A$-$S_p$-$T_{exp}/T_p$ & \multirow{2}{*}{99.6} & \multirow{2}{*}{95.2}  & \multirow{2}{*}{88.5} & \multirow{2}{*}{93.0} & \multirow{2}{*}{94.9} \\
KNN (this work) &  &  &  &  &  &  &  \\ \hline
$B_T$-$N_P$-$T_P$-$V_P$-$N_{\alpha p}$-$T_{exp}/T_P$-$S_p$-$M_f$ & \multirow{2}{*}{99.2} & \multirow{2}{*}{91.1} & \multirow{2}{*}{83.8} & \multirow{2}{*}{92.9}  & \multirow{2}{*}{92.8} \\
KNN (this work) &  &  &  &  &  &  &  \\
\hline
\hline
\end{tabular}
\label{compare}
\end{table}

Table \ref{compare} gives the comparison of the performances of various categorization schemes. The O$^{7+}$/O$^{6+}$-$V_p$ scheme proposed by \citet{Zhao et al 2009} can not distinguish SBP and SRRP, and does not work well for identifying EJECT. The accuracy is only 63.5\%. \citet{Xu and Borovsky 2015} proposed the $S_p$-$V_A$-$T_{exp}/T_p$ scheme, which has a significant improvement on identifying EJECT and increases the accuracy to 87.5\%. In addition, such a scheme can also distinguish SBP and SRRP, with an accuracy $\sim$ 70\%. Note that, the classification accuracies obtained by \citet{Xu and Borovsky 2015} are quite comparable to those obtained by KNN classifier. By taking the advantage of machine learning on classification in multi-dimensional parameter space, we apply an eight-dimensional scheme, the $B_T$-$N_P$-$T_P$-$V_P$-$N_{\alpha p}$-$T_{exp}/T_P$-$S_p$-$M_f$ scheme, on KNN classifier, and obtain significant improvements in classification accuracies. The improvements of accuracy for identifying CHOP, SBP, SRRP, and EJECT is 2.3\%, 21.2\%, 11.8\%, and 5.4\%, respectively. For the 4-type solar wind classification, the overall accuracy has an improvement of 9.6\%. It should be mentioned that, the feature space has been optimized only for the KNN approach. For other classifiers with some other parameter scheme used, the accuracies could be improved. \citet{Camporeale et al 2017} proposed a classification scheme based on Gaussian Process classifier. By using the $V_p$-$\sigma_T$-$SSN$-$F10.7$-$V_A$-$S_p$-$T_{exp}/T_p$ scheme ($\sigma_T$ is the standard deviation of proton temperature, $SSN$ is the sunspot number, and $F10.7$ is the solar radio flux at 10.7 cm), they concluded that they have obtained ``excellent'' classification accuracies, which are better than 96\% for all the four types of solar wind. Note that, 25\% of the database is used for training in \citep{Camporeale et al 2017}, but the ratio is 75\% in our study. If the same parameter scheme was performed on KNN classifier, the overall classification accuracy has a slight decrease of 3.3\%, although the accuracy for SRRP has a decrease of 9.0\%, as listed in Table \ref{compare}. This indicates that the performance of KNN classifier is close to, or not far worse than, that of Gaussian Process classifier used by \citet{Camporeale et al 2017}. 

We apply the trained KNN classifier to classify the OMNI data set from 1963 to 2017. The probabilities of CHOP, SBP, SRRP, EJECT are obtained. As mentioned before, the threshold of probability is chosen to be 0.3-0.5 to obtain optimal TPR and FPR. The event with the maximum of probability less than the threshold is defined as an ``undecided'' event. If the threshold is chosen to be 0.3, the percentage of ``undecided'' events is 0.02\%. And if the threshold is chosen to be 0.5, the percentage of ``undecided'' events is less than 2.2\%. For comparison, the percentage of ``undecided'' events is 0.2\% and 7.5\% in \citet{Camporeale et al 2017}, indicating that the possibility of ``undecided'' solar wind type is larger than our approach.

\begin{figure}[htbp]
\centering
\noindent\includegraphics[width=32pc]{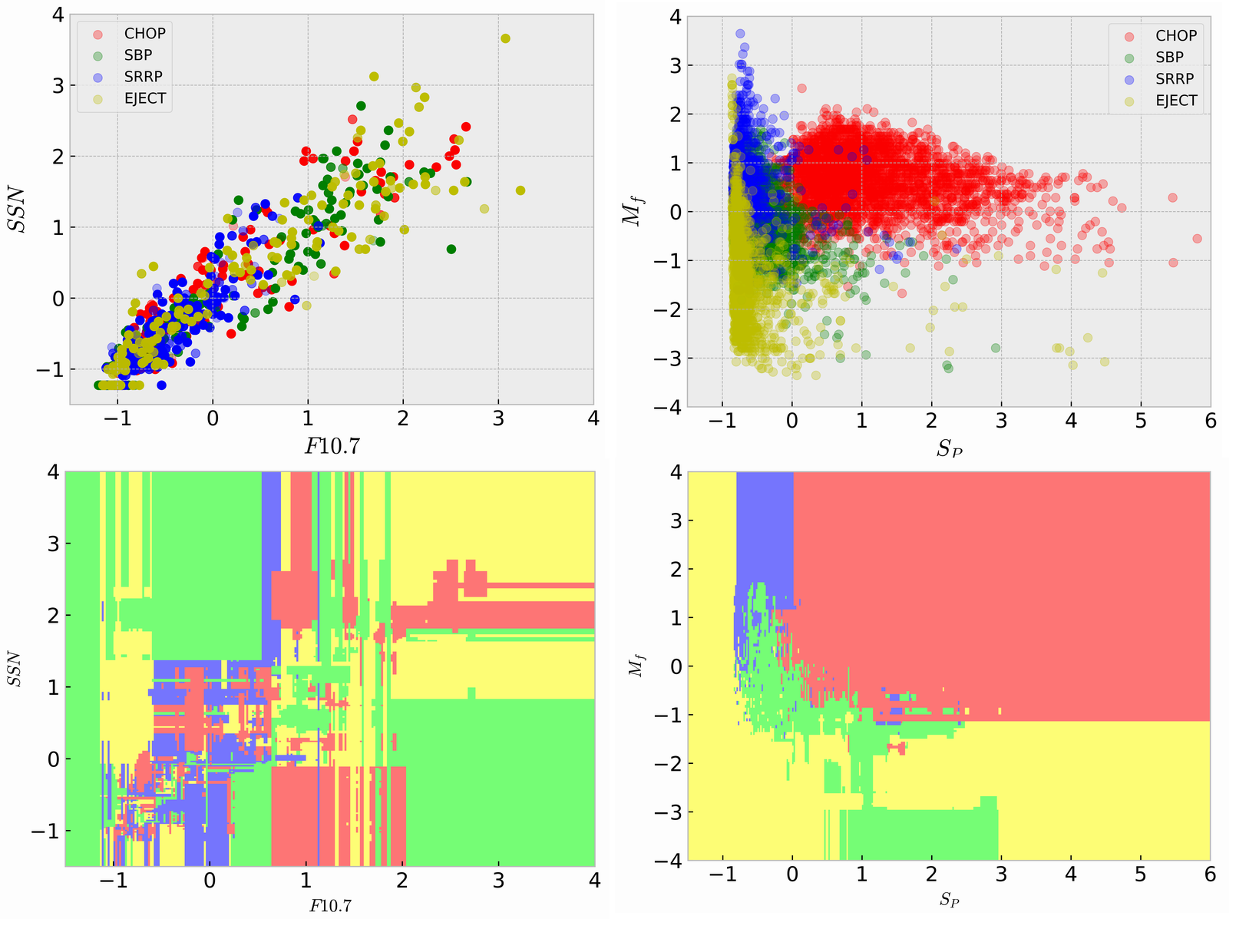}
\caption{Top: Distribution of reference solar wind in the plot of $SSN$ vs. $F10.7$and $M_f$ vs. $S_p$. Bottom: Corresponding decision boundaries for each solar wind category. The overall accuracy given by KNN classifier under the $SSN$-$F10.7$ scheme is 98.5\%, however, the accuracy is 79.2\% under the $M_f$-$S_p$ scheme.}
\label{comp}
\end{figure}

\section{Discussion}
\subsection{daily sampled parameters are not recommended for hourly solar wind classification}
\citet{Camporeale et al 2017} used the mixture of hourly averaged solar wind parameters and daily sampled parameters to obtain ``excellent'' classification accuracies, however, it is not recommended in this study. Firstly, the time resolution of $SSN$ and $F10.7$ used by \citet{Camporeale et al 2017} is one day, which does not match the temporal resolution of reference solar wind data sets and other five solar wind parameters, which are one hour. \citet{Camporeale et al 2017} did not demonstrate the reasonableness of such an approach. Secondly, $SSN$ and $F10.7$ are found to be questionable in solar wind classification. Taking the $SSN$-$F10.7$ scheme for example, the overall accuracy obtained by KNN classifier is 98.5\%, and is 99.5\%, 98.9\%, 98.1\%, and 96.6\%, for CHOP, SBP, SRRP, and EJECT, respectively, which is even better than the results of \citet{Camporeale et al 2017}. However, it does not indicate that $SSN$-$F10.7$ scheme is a better choice in solar wind classification. As shown in Figure \ref{comp}, it is quite difficult to distinguish CHOP, SBP, SRRP, and EJECT from each other in the plot of $SSN$ vs. $F10.7$. At the same time, the corresponding decision boundaries for each solar wind category are too complicate to eliminate the concerns on the probability of over-fitting problem. One plausible reason is that there are only 479 independent data points in the plot of $SSN$ vs. $F10.7$ for 8625 events. The ratio of independent data ($<$ 6\%) is much less than the ratio of training dataset (75\% in our work and 10\%-25\% in \citet{Camporeale et al 2017}), which means that all the independent data are used both for training and for testing. In other word, the cross validation does not work at all, which may result in a lager risk of over-fitting problem. For comparison, the distribution of reference solar wind in the plot of $M_f$ vs. $S_p$ is also shown in Figure \ref{comp}. Although the overall accuracy given by KNN classifier is 79.2\%, much lower than that for $SSN$-$F10.7$ scheme, it is still possible to generally distinguish the distribution of CHOP, SBP, SRRP, and EJECT from each other, except a few overlaps. And the decision boundaries are more likely to represent a regularized classification. Thirdly, \citet{Camporeale et al 2017} did not explain that the classification performance significantly increases when using both $SSN$ and $F10.7$ with respect to the case when one of the two is left out, although $SSN$ and $F10.7$ are strongly correlated. Thus, it is strongly suggested to use the solar wind parameters with the same time resolutions as the reference solar wind data sets when training the classifier.

\subsection{Composition information in solar wind classification}
In the previous classification schemes in two- or three- dimensional parameter space, solar wind composition measurement indeed plays an important role in solar wind classification. However, it is still difficult to conclude that the composition measurement is thus indispensable. To show the importance of composition information in solar wind classification, we have accessed the 1-hr composition data (C6+/C5+, O7+/O6+) from ACE satellite during 1998-2011. During this time interval, the reference solar wind data sets with data gap removed are 8021 hours: CHOP (2881 hours), SBOP (2215 hours), SRRP (1694 hours), Ejecta (1231 hours). Compared to the data sets without composition information, the Ejecta data reduced from 1835 hours to 1231 hours, and the CHOP, SBOP, SRRP data are the same. The overall classification accuracy by solely using C$^{6+}$/C$^{5+}$ or O$^{7+}$/O$^{6+}$ is 51.0\% and 65.9\%, which is less or comparable to the performance, 66.7\%, when Sp is used solely.

The comparison of classification results with/without composition information is shown in Table \ref{composition}. It is clear that the classification results indeed have some minor improvements, especially when O$^{7+}$/O$^{6+}$ information is considered. But the improvements are not much significant, only 1.5\%. Considering that most of solar wind satellites, for example, the recent Parker Solar Probe, do not have composition instrument, it is suggested that solar wind classification scheme without composition information is still useful.

\begin{table}[htbp]
\renewcommand{\thetable}{\arabic{table}}
\caption{Comparison of solar wind classification with/without composition information}
\centering
\begin{tabular}{c c c c c c c c}
\hline
\hline
          & CHOP & SBP & SRRP & EJECT & 4-type & HKSS\\
\hline
   Without Composition  & 99.3 & 91.4 & 85.1 & 92.5 & 93.1 & 0.903 \\
   $C^{6+}/C^{5+}$ & 99.3 & 92.5 & 85.6 & 92.6 & 93.5 & 0.909 \\
   O$^{7+}$/O$^{6+}$ & 99.4 & 93.0 & 89.0 & 94.1 & 94.6 & 0.925 \\
   C$^{6+}$/C$^{5+}$ \& O$^{7+}$/O$^{6+}$ & 99.4 & 93.2 & 87.3 & 93.1 & 94.3 & 0.920 \\
\hline
\hline
\end{tabular}
\label{composition}
\end{table}

\subsection{Importance of the accuracy of reference solar wind}
The reference solar wind with known types is very important for supervised machine learning. In this study, the reference solar wind data comes from the work based on human experiences, which may have some uncertainties, especially at the boundaries of events. A natural thought is that the center part of an event has the highest probability to be correctly labeled. For practice, if 3-hr data points at both boundaries were deleted for each EJECT event, the classification accuracy of EJECT should have an improvement of 2.2\%. Thus, the further improvement of classification accuracy by machine learning is limited by the uncertainties of the reference solar wind data set.

\section{Application in Space Weather Early Warning} \label{sec:application}

The information of solar wind origin may be helpful for the early warnings of space weather. Firstly, the solar wind category is useful for the risk evaluation of a predicted geomagnetic storm. \citet{Turner et al 2009} showed that the storm intensity and occurrence rate of intense storm (Dst minimum $<$ -100 nT) for ICME-driven storms are larger than that for CIR-driven storms. The average Dst minimum during a CIR-driven storm is $\sim$ -74 nT, and the occurrence rate of intense storms is only 13\%, however, these two values are -128 nT and 57\% for ICME-driven storms, respectively. Besides, all superstorms, with Dst minimum $<$ -300 nT and midday magnetopause shifting earthward of geosynchronous orbit \citep{Li et al 2010}, are associated with ICMEs. Secondly, the classification of CHOP and EJECT is also helpful for the risk evaluation of surface charging of geosynchronous spacecrafts. \citep{Borovsky and Denton 2006,Denton et al 2006} found that the magnitude of spacecraft potential is, on average, significantly elevated for CIR-driven storms than during ICME-driven storms. Thirdly, \citet{McGranaghan et al 2014} showed that SBP and SRRP produce forecastable changes in thermospheric density.

\citet{Gonzalez and Tsurutani 1987} suggested that storm intensity depends on the intensity of southward interplanetary $B_Z$ and the threshold for intense storms is summarized to be -10 nT. \citet{Echer et al 2008} later found that storm intensity depends on the solar wind electric field $E_Y$ and the threshold for intense storms is summarized to be 5 mV/m. If $B_Z$ $\le$ -10 nT and $E_Y \geq$ 5 mV/m are observed in the solar wind at L1 point, a magnetic storm is likely to occur in the next several hours. With the information of solar wind type obtained, more details of the geoeffectiveness will be inferred. Table \ref{app} gives two examples. For the first case, $B_Z$ is observed to be -11.2 nT on 00:00 Feb-27-2003, moreover, the corresponding $E_Y$ is observed to be 5.03 mV/m. Based on our classification algorithm, the solar wind plasma is categorized to be SBP, indicating a possible CIR-driven storm. \citet{Borovsky and Denton 2013} indeed identified that event as a pseudostreamer CIR. Thus, the impending storm will be predicted to be a moderate storm with a big probability, at the same time, the risk of dangerous spacecraft surface charing is predicted to be high. As a validation, the real occurred storm is identified to be a moderate storm, with the Dst minimum of -60 nT. Besides, the magnitude of spacecraft potential ($\Phi$) in geosynchronous orbit during this storm is close to 4000 V. For the second case, similar $B_Z$ and $E_Y$ are observed to be -10.8 nT and 5.08 mV/m on 00:00 Nov-08-1998. Unlikely, the solar wind plasma is categorized to be EJECT for this case, (which is later identified as an ICME by Richardson and Cane later, http://www.srl.caltech.edu/ACE/ASC/DATA/level3/icmetable2.htm), indicating a possible ICME-driven storm. Thus, the impending storm will likely be an intense storm, however, the risk of spacecraft surface charging is predicted to be relative low. In fact, the following storm has an intensity of -149 nT and the magnitude of spacecraft potential during this storm is no more than 900 V.

\begin{table}[htbp]
\renewcommand{\thetable}{\arabic{table}}
\caption{Application of the information of solar wind origin in improving space weather forecast.}
\centering
\begin{tabular}{c c c c c c c}
\hline
\hline
 Time & $B_Z$ & $E_Y$ & Type & Forecast & $Dst_{min}$ & $\Phi$ $^{a}$ \\
\hline
 Feb-27-2003 & \multirow{2}{*}{-11.2} & \multirow{2}{*}{5.03} & \multirow{2}{*}{SBP} & Moderate CIR-storm & \multirow{2}{*}{-60} & \multirow{2}{*}{4000}\\ 
 00:00 UT &  &  &  & high charging risk &  & \\ \hline

  Nov-08-1998 & \multirow{2}{*}{-10.8} & \multirow{2}{*}{5.08} & \multirow{2}{*}{EJECT} & Intense ICME-storm & \multirow{2}{*}{-149} & \multirow{2}{*}{900}\\ 
  00:00 UT &  &  &  & low charging risk &  & \\ \hline
\hline
\multicolumn{2}{l}{$^{a}$ Data from the LANL/MPA instrument.}
\end{tabular}
\label{app}
\end{table}

At present, we use the in-situ observation at L1 point to classify the solar wind, and can make a space weather early warning by half an hour or more. There could be more utility for the present classification scheme if a solar wind monitor is placed at L5. Besides, we are still working on improving the time advance of solar wind classification by using the observations on the Sun's surface.

\section{Summary} \label{sec:summary}

Solar wind categorization is conducive to understanding the solar wind origin and physical processes ongoing at the Sun. Facing a great deal of spacecraft observations, manual classification based on rich experiences is prohibitive in terms of time and is challenged. Automatic classification methods are quite needed. Recently, with rapid developments in the field of artificial intelligence, the classification by machine learning is becoming more and more popular and powerful in big-volume data analysis, and furthermore, its performance is improving as well.

In this study, 10 popular supervised machine learning models, $k-$nearest neighbors (KNN), linear support vector machines (LSVM), SVM with a kernel of Gaussian radial basis function (RBFSVM), Decision Trees (DT), Random Forests (RF), Adaptive Boosting (AdaBoost), Neural Network (NN), Gaussian Naive Bayes (GNB), Quadratic Discriminant Analysis (QDA), and eXtreme Gradient Boosting (XGBoost), are used to classify the solar wind at 1 AU into four plasma types: coronal-hole-origin plasma, streamer-belt-origin plasma, sector-reversal-region plasma, and ejecta.

A total of 13 parameters, each with 1-hr temporal resolution, are used for training the classifiers and searching for the best variable scheme. These parameters are the magnetic field intensity $B_T$, the proton number density $N_P$, the proton temperature $T_P$, the solar wind speed $V_P$, the proton-specific entropy $S_p$, the Alfv\'en speed $V_A$, the ratio of velocity-dependent expected proton temperature and proton temperature $T_{exp}/T_P$, the number density ratio of proton and alpha $N_{\alpha p}$, the dynamic pressure $P_d$, the solar wind electric field $E_y$, the plasma beta value $\beta$, the Alfv\'en Mach number $M_A$, and the fast magneto-sonic Mach number $M_f$. Note that, all the parameters can be obtained or derived from the typical solar wind observations. No composition measurements are needed, allowing our algorithm to be applied to most solar wind spacecraft data. 

By exhaustive enumeration, an eight-dimensional scheme ($B_T$, $N_P$, $T_P$, $V_P$, $N_{\alpha p}$, $T_{exp}/T_P$, $S_p$, and $M_f$) is found to obtain the highest classification accuracy among all the 8191 combinations of the above 13 parameters. Among the 10 popular classifiers, the KNN classifier obtains an accuracy of 92.8\%. It significantly improves the accuracy by 9.6\% over existing manual schemes. In addition, small-scale flux rope events may also be able to be identified based on our method, though further validation is needed. Besides, two application examples of solar wind classification are given, indicating that it is helpful for the risk evaluation of predicted magnetic storms and surface charging of geosynchronous spacecrafts. 

This work emphasizes the classification technique itself rather than the science of the solar wind origin. In the future, with new solar wind types and corresponding ideal events are proposed in the community, our machine learning approach will be updated accordingly and more efforts are needed to bring up some new understandings to the science of the solar wind origin.

\acknowledgments

We thank the OMNI database for the use of solar wind data, which is accessible in ftp://spdf.gsfc.nasa.gov/pub/data/omni/low\_res\_omni/. We also thank the scikit-learn and XGBoost toolkits written in Python, which provide the classification classifiers used here and can be found in https://scikit-learn.org/stable/install.html. The reference solar wind data for training and testing used in this study, and the final classified solar wind can be accessed in http://www.spaceweather.ac.cn/\%7Ehli/research.html. This work was supported by Strategic Priority Research Program of Chinese Academy of Sciences (Grant No. XDA17010301), NNSFC grants 41874203, 41574169, 41574159, 41731070, Young Elite Scientists Sponsorship Program by CAST, 2016QNRC001, and grants from Chinese Academy of Sciences (QYZDJ-SSW-JSC028, XDA15052500). H. Li was also supported by the Youth Innovation Promotion Association of the Chinese Academy of Sciences, and in part by the Specialized Research Fund for State Key Laboratories of China.

%




\listofchanges

\end{document}